\documentclass[sigconf]{acmart}

\usepackage{booktabs} 
\usepackage{url}
\setcopyright{rightsretained}

\acmDOI{}

\acmISBN{}

\author{Farhan Khawar}
\affiliation{\institution{The Hong Kong University of Science and Technology}}
\email{fkhawar@cse.ust.hk}

\author{Nevin L. Zhang}
\affiliation{\institution{The Hong Kong University of Science and Technology}}
\email{lzhang@cse.ust.hk}
\begin{document}

\title{Using Taste Groups for Collaborative Filtering}


\begin{abstract}
Implicit feedback is the simplest form of user feedback that can be used for item recommendation. It is easy to collect and domain independent. However, there is a lack of negative examples. Existing works circumvent this problem by making various assumptions regarding the unconsumed items, which fail to hold when the user did not consume an item because she was unaware of it. In this paper, we propose as a novel method for addressing the lack of negative examples in implicit feedback. The motivation is that if there is a large group of users who share the same taste and none of them consumed an item, then it is highly likely that the item is irrelevant to this taste. We use Hierarchical Latent Tree Analysis(HLTA) to identify taste-based user groups and make recommendations for a user based on her memberships in the groups.
 \end{abstract}

%
%







\keywords{Collaborative filtering; Implicit Feedback; OCCF }
\maketitle

\section{Introduction}
A key issue with implicit feedback is how to deal with the lack of negative examples. We are unsure whether the user didn't consume an item because she didn't like it or because she never saw it.
In this paper, we propose a novel method for addressing this issue. We start by identifying user groups with the same tastes. By a \emph{taste} we mean the tendency to
consume a certain collection of items such as comedy movies, pop songs, or spicy food. Those taste-based groups give us a nice way to deal with the lack of negative examples. While it is not justifiable to assume that non-consumption means disinterest for an individual user, it is relatively more reasonable to make that assumption for a group of users with the same taste: If  many users share a taste and none of them have consumed an item before, then it is likely that the group is not interested in the item.

We use HLTA \cite{chen2017latent}
to identify taste-based user groups.  When applied to implicit feedback data, HLTA obtains a hierarchy of binary latent variables by: (1) Identifying item co-consumption patterns (groups of items that tend to be consumed by the same customers, not necessarily at the same time) and introducing a latent variable for each pattern; (2) Identifying co-occurrence patterns of those patterns and introducing a latent variable; (3) Repeating step 2 until termination.  Each of the latent variables identifies a soft \emph{group of users}, just as the concept ``intelligence" denotes a class of people. 
To make recommendations, we choose the user groups from a certain level of the hierarchy and characterize each group by aggregating recent behaviors of its members. For a particular user, we perform inference on the learned model to determine her memberships in the groups, and predict her preferences  by combining her memberships in the groups and the group characteristics.
\section{Related Work}\label{related}

\begin{figure}
\begin{center}
\includegraphics[width=7cm, height=4cm]{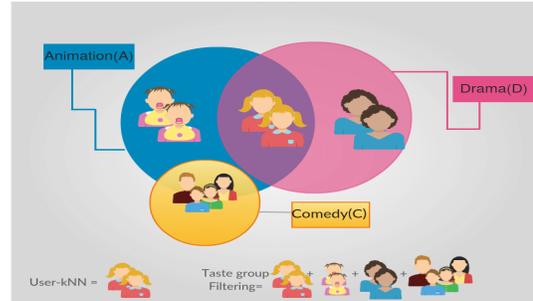}
\end{center}
\vspace{-0.5cm}
\caption{ Groups of people used by user-kNN and taste-group filtering to recommend items for a user who has three tastes in videos: Animation, Comedy and Drama. Each circle in the Venn diagram represents a group of people who have a taste for the video genre indicated. User-kNN uses the intersection of the three groups while taste-group uses their union. }
\vspace{-0.5cm}
\label{CoF}
\end{figure}

Taste-group filtering is similar to user-kNN  in that they both predict a user's preferences based on past behaviors of similar users. There are two important differences. First, when a user belongs to multiple taste groups, as is usually the case, our method uses information from all the users in those groups, while user-kNN uses information only from the users who are in all groups. To put it another way, our method uses the union of the groups, while user-kNN uses their intersection. This is illustrated in Figure \ref{CoF}. Second, user-kNN is not model-based whereas our method is.  More specifically, the taste groups are the latent factors. An item is characterized by the frequencies it was consumed by members of the groups, and a user is characterized by her memberships in the groups. Moreover, in comparison with  matrix factorization, our latent factors are more interpretable. They also an additional flexibility of using recent behaviors of group members, instead of their entire consumption histories, when predicting future behavior of the group.

The assumptions behind all other existing methods e.g., WRMF\cite{Hu:2008:CFI:1510528.1511352}, BPRMF\cite{Rendle:2009:BBP:1795114.1795167}, SLIM\cite{ning2011slim} etc. are regarding the preferences of individual users. They fail to hold when a user $u$ would have liked an item $i$ but did not consume it only because she was unaware of it. In that case, it is incorrect to assume user $u$ is not interested in item $i$, even with low confidence; it would be a mistake if the pair $(u, i)$ is sampled as a negative example; and it is wrong to assume $u$ prefers all her consumed items to item $i$.
In contrast, the assumption behind out method is about the preferences of groups of users. If a group is large enough, it is relatively safe to assume that most of the items have come to the attention of at least one of the group members, and hence relatively reasonable to assume that the items not consumed by any group members are not of interest to the group.

\section{Basics of Latent Tree Models} \label{LTM}
\begin{figure}[]
	\centering
	\includegraphics[scale=0.12]{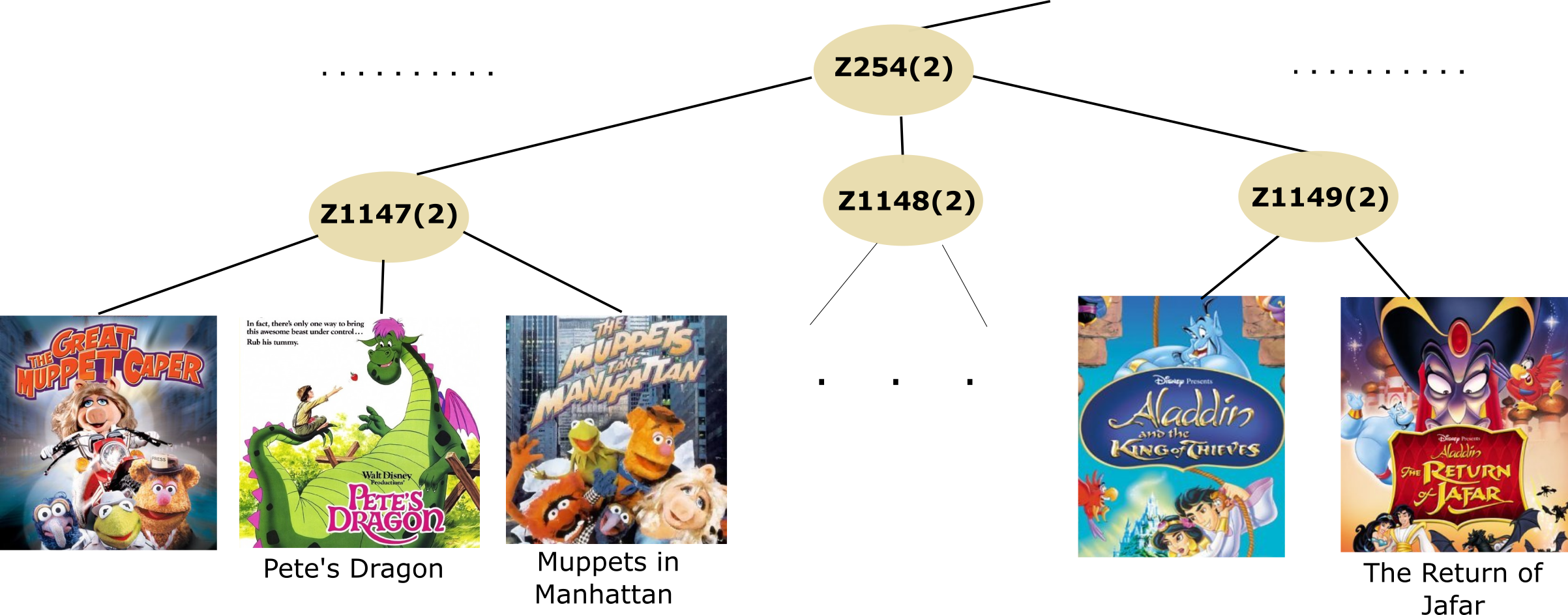}
	\caption{Part of the LTM obtained from Movielens1M dataset. The level-1 latent variables reveal  tastes for various subsets of items. Latent variables at higher levels identify more broad tastes.}
	\label{fig:lastfm1k-5core-train70reduced}
\label{fig:Movielens1M}
\end{figure}

A {\em latent tree model (LTM)} is a tree-structured Bayesian network, where the leaf nodes represent observed variables and the internal nodes represent latent variables. An example is shown in Figure \ref{fig:Movielens1M}. All variables are assumed to be binary. The model parameters include a marginal distribution for the root and a conditional distribution for each of the other nodes given its parent. The product of the distributions defines a joint distribution over all the variables. To learn an LTM, one needs to determine: (1) the number of latent variables, (2) the connections among all the variables, and (3) the probability distributions from the data.

\section{Taste-based Filtering}\label{CF}

Suppose we have learned an LTM $m$ from implicit feedback data and suppose there
are $K$ latent variables on the $l$-th level of the model, each with two states $s0$ and $s1$. Denote the latent variables as  $Z_{l1}$, \ldots, $Z_{lK}$. They give us $K$ taste-based user groups
$Z_{l1}=s1$, \ldots, $Z_{lK}=s1$, which will sometimes be denoted as
$G_1$, \ldots, $G_K$ for simplicity. In this section, we show how these taste groups can be used for item recommendation.

\subsection{Taste Group Characterization}
A natural way to characterize a user group is to aggregate past behaviors of the group members. The issue is somewhat complex for us because our user groups are soft clusters. Let $\mathbb{I}(i|u, {\mathcal D})$ be the indicator function which takes value 1 if user $u$ has consumed item $i$ before, according to the dataset ${\mathcal D}$, and 0 otherwise. We determine the preference of a taste group $G_k$ (i.e., $Z_{lk}=s1$) on an item $i$  as follows:

\vspace{-4.5mm}
\begin{eqnarray}
p(i|G_k, {\mathcal D}) = \frac{\sum_{u} \mathbb{I}(i|u, {\mathcal D})
P(Z_{lk}=s1|u, m)}{\sum_{u} P(Z_{lk}=s1|u, m)},
\label{eq.group}
\end{eqnarray}
\noindent where $P(Z_{lK}=s1|u, m)$ is the probability of user $u$ being in
the soft cluster $Z_{lk}=s1$.

Note that $p(i|G_k, {\mathcal D})=0$ if no users in $G_k$ have consumed the item $i$ before.
In other words, we assume that a group is not interested in an item if none of the group members have consumed the item before. 
\subsection{User Characterization and Recommendation}

A user $u$ is characterized using her memberships in the $K$ clusters, i.e., $	{\bf u}= (P(Z_{l1}=s1|u, m), \ldots, P(Z_{lK}=s1|u, m)).$ The score $\hat{r}_{ui}$ for a user-item pair $(u, i)$ is computed by combining the taste group characterizations and the memberships of $u$ in those groups: $\hat{r}_{ui} = \sum_{k=1}^K p(i|G_k, \mathcal{D}_H) P(Z_{lk}=s1|u, m).$

\section{Results and Future Work}
\begin{figure}[]
	\centering
	\includegraphics[scale=0.45]{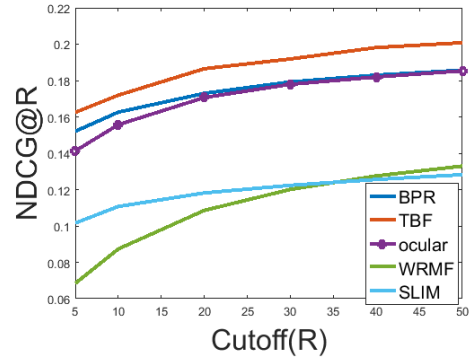}
	\caption{NDCG results at different cutoff points.}
	\label{fig:ndcg}

\end{figure}

We performed experiments on the Ta-feng supermarket dataset which contains binary purchase events. The dataset was split in train, validation and test sets based on the time-stamps with a ratio of 70\%,15\% and 15\% respectively. The NDCG@R results are show in Figure \ref{fig:ndcg} and AUC in Table \ref{tab:AUC}. As can be seen Taste-based filtering (TBF) achieves better performance compared to the baselines.

It remains to be verified that the performance gain is due to the taste-group assumption. 
The potential explainability of the method is yet to be explored.

\begin{table}[]
	\centering
	\vspace{-3mm}
	\caption{The AUC for each recommender is shown. TBF outperforms other methods and BPR comes second as expected.}
	\label{tab:AUC}
	\begin{tabular}{llllll}
		\hline
		&\textbf{BPRMF}   & \textbf{WRMF}    & \textbf{TBF}     & \textbf{Ocular}  &\textbf{ SLIM}            \\
		\hline
		Ta-feng & 0.74977 & 0.71316 & \textbf{0.7793}  & 0.63653 & 0.62949 \\
		\hline      
	\end{tabular}
\end{table}	

Research on this article was supported by Hong Kong Research Grants Council under grant 16202118.

\bibliographystyle{ACM-Reference-Format}
\bibliography{CoF} 

\end{document}